\documentstyle[11pt]{article}

\title{An Extended Technicolor Model}
\author{Thomas Appelquist \\
Department of Physics, Yale University, New Haven CT 06511\\
and\\
 John Terning\\
Department of Physics, Boston University\\
590 Commonwealth Ave., Boston MA  02215}
\begin{document}
\setlength{\baselineskip}{24pt}
\maketitle
\begin{picture}(0,0)(0,0)
\put(295,285){YCTP-P21-93}
\put(295,270){BUHEP-93-23}
\end{picture}
\vspace{-24pt}

\begin{abstract}An extended
technicolor model is constructed.
Quark and lepton masses, spontaneous CP violation, and precision electroweak
measurements are discussed.  Dynamical symmetry breaking is analyzed using the
concept of the BIG MAC.
\end{abstract}

\section{Introduction}
Recent work \cite{walking,strongETC,KMEN,Sundrum,revenge}
on technicolor (TC) models
indicates that it may be possible to
describe the observed particle mass spectrum, while avoiding flavor changing
neutral currents (FCNC's) and satisfying precision electroweak tests.  That is,
using a phenomenologically acceptable TC gauge group and
technifermion count, and representing the extended
technicolor (ETC) interactions
by four-fermion interactions with arbitrary mass scales and
arbitrary
couplings for each of the ordinary fermions, one can produce the
entire observed
range of fermion masses, up to well over one hundred GeV for the $t$
quark, without excessive fine tuning of parameters and without any
phenomenological disasters.
Though this exercise is interesting as a sort of existence
proof, it uses as many parameters
as observables, so it is difficult to be sure if success is the
result of having identified the correct physics.

To do better one must construct a model that explains the world rather
than just
describes it, i.e. a model with fewer parameters than the standard
model.  It is our purpose
here to construct a plausible ETC model.  After reviewing the constraints
that must be satisfied, we
will present the model.  We then conclude with a discussion
of quark and lepton masses, precision electroweak tests,  and CP
violation.  One additional prediction for new phenomena will also be described.

\section {Ingredients for Model Building}
There are several ingredients that should be incorporated
into a realistic
ETC model.  First of all, more than one ETC scale is expected. The absence of
FCNC's (inferred from  $K-\overline{K}$
mass splitting) requires the mass of the ETC bosons that
connect to the
$s$ quark  to be at least\footnote{Assuming a coupling of order 1, and the
absence of a ``TechniGIM" mechanism \cite{Randall}.}  about  $\Lambda_{FCNC}
=1000$ TeV $\cos \theta\, \sin\theta$, where $\theta$ is a model dependent
mixing angle \cite{EL,Ellis}.   For example, taking $\theta$ to be equal to the
Cabibbo angle, we find $\Lambda_{FCNC} = 200$ TeV.   In order to have such a
high ETC scale associated with the $s$, and still
produce the correct mass, one may have to invoke walking \cite{walking}.
Also, to obtain a $t$ quark mass above 100 GeV without excessive fine tuning,
it turns
out that the ETC scale relevant to $t$ mass generation should be at most  about
10 TeV \cite{strongETC}.  Such arguments, coupled with the observation of the
hierarchy of family masses, suggest  three different ETC scales, one for each
family.  In this paper we will take these scales to be roughly 10, 100, and
1000 TeV.   With a reasonable running of gauge
couplings, these scales can arise naturally via self-breaking gauge
interactions, and may thus
afford us with a natural explanation of the family mass hierarchy.

A realistic ETC model must also survive
precision electroweak tests \cite{rho,PT}. It must produce a
large $t$-$b$ mass splitting, while
keeping the radiative electroweak correction parameter
$\Delta\rho_*\equiv \alpha T$ less than about  $ 0.5\%$.
The radiative electroweak correction parameter $S$  can also be
worrisome \cite{PT}. Experiments
seem to be finding $S$ to be very small or even negative,
whereas QCD-like
TC models give positive contributions to $S$ (which grow with the
number of technicolors, $N_{TC}$).  Of course, QCD-like
TC models may already be ruled out since they lead to large FCNC's, and
furthermore it is
difficult to reliably estimate $S$ in TC models with non-QCD-like
dynamics \cite{CHS}.
Nevertheless, the constraint on the $S$ parameter seems to suggest that
$N_{TC}$ should be kept as small as possible.

An important constraint on ETC model building was originally
elucidated by Eichten and Lane \cite{EL},
who showed that the absence of a visible axion implies a limit on
the number of spontaneously broken global U(1)
symmetries\footnote{Another possibility is that the axion is made very
heavy by QCD, see ref. \cite{HP}.}, and hence a limit on the number of
irreducible
representations of the ETC gauge group.
This points to some form of quark-lepton unification (such as
Pati-Salam unification \cite{PS}), in
ETC models.

Also, to avoid a plethora of  massless, non-Abelian, Nambu-Goldstone bosons,
a realistic ETC model should not have any
exact, spontaneously broken,  non-Abelian global symmetries.  Thus there
should not be repeated representations of the ETC gauge group.

An ETC model must also explain why neutrinos are special. The fact that only
extremely light left-handed neutrinos are seen
in nature is one of the most puzzling features of the quark-lepton mass
spectrum. It poses special problems for
ETC model builders, since it is difficult to construct ETC
models without right-handed neutrinos.
With right-handed neutrinos present in the model, there are at least two
simple explanations available for small neutrino masses:
an implementation of
the usual seesaw mechanism \cite{GMRS,models},
or the possibility
that the technifermion masses do not feed down directly to
the neutrinos.   The latter possibility was  suggested long ago by
Sikivie, Susskind, Voloshin, and Zakharov (SSVZ) \cite{SSVZ}. The
model to be discussed in this paper will utilize this mechanism.

SSVZ considered an $SU(3)_{ETC}$ gauge group (which will
appear in our model below 100 TeV), where a {\bf 3} of $SU(3)_{ETC}$
corresponds to two technifermions and one third-generation fermion.
$SU(3)_{ETC}$ will be broken to $SU(2)_{TC}$ by another strong gauge
interaction, referred to here as hypercolor (HC).  The idea of SSVZ is to
put leptons in unusual  ETC representations.   The left-handed leptons are
placed in {\bf 3}'s of $SU(3)_{ETC}$; the
charge-conjugated, right-handed charged-leptons in
${\bf \overline{3}}$'s; and the charge-conjugated, right-handed neutrinos
in {\bf 3}'s.  When $SU(3)_{ETC}$ breaks, all the technileptons are
in equivalent $SU(2)_{TC}$ representations, but the ETC interactions of the
$\nu_\tau$
are different  from those of  the $\tau$.  In fact,
it can be shown that to leading order in ETC exchange, the $\nu_\tau$
does not receive a mass. ETC interactions must of course
be extended beyond $SU(3)_{ETC}$. The $\nu_\tau$ may receive a mass in higher
orders in these interactions, but, as we will discuss later, in the model to be
presented, at two loops, the $\nu_{\tau L}$ will remain massless.

Our model will also ensure that the muon neutrino ($\nu_\mu$)  mass vanishes to
a sufficiently high order in perturbation theory so as to satisfy the
experimental constraint on its mass.
This will arise through a simple extension of the SSVZ mechanism to the second
generation.

Whether or not the SSVZ mechanism  is extended to the first generation, there
would be other contributions to the
$\nu_e$ mass that are much too large.  If quarks and leptons
are unified (as they must be in a realistic ETC model),  then masses can feed
down to
$\nu_e$ from ordinary fermions as well as technifermions.
For example, consider  a Pati-Salam \cite{PS}
unification scheme. The $\nu_e$ is placed in the same
representation as the $u$ quark, and  there
is a diagram that feeds the $u$ quark mass down to the $\nu_e$ through
the exchange of a heavy Pati-Salam gauge boson\footnote{In a self-consistent
calculation one should also include the $\nu_e$ self-energy coming from a
single Pati-Salam gauge boson exchange.}.   A standard calculation (for
simplicity taking the Pati-Salam breaking scale, which provides the cutoff for
the calculation, to be equal to the ETC scale of the first generation) then
gives:
\begin{equation}
m_{\nu_e} \approx
 {{9 \, \alpha_{PS}}\over{8 \pi}}  m_u  ~.
\end{equation}
This gives a mass for the $\nu_e$  on the order of a fraction of
an MeV, far above the experimental bound.  In order to avoid such
a disaster, the right-handed neutrino that is unified with the
right-handed $u$ quark must get either a large Majorana mass with
itself, or a large Dirac mass with another $SU(2)_L$ singlet
neutrino.  The model presented here will employ the later possibility, and as a
result there will be no right-handed neutrino in the first family.

 Another problem that ETC models must face is intrafamily mass
splittings. The most striking such splitting, and
the most difficult to account for in models with a family of
technifermions, is the $t$-$\tau$ splitting.  One possible
solution is that this splitting comes from QCD effects
\cite{color,color2}.  It is possible for small perturbations (like QCD) to
have large effects in models where the (strong) ETC coupling is near
critical \cite{strongETC}.  One calculation \cite{color2} found that
this effect could give a quark mass up to two orders of magnitude
larger than that of the corresponding lepton.  We will rely on the efficacy of
this
mechanism in our model.

\section{A Realistic ETC Model}

We now construct an ETC model, using the smallest possible TC group:
$SU(2)_{TC}$.  One family of technifermions will be included,
since this allows for the smallest possible ETC gauge group.
We require that:

1) there are no exact non-Abelian global symmetries,

2) quarks and leptons are unified so as to avoid a visible techniaxion,

3) fermions are only allowed to be singlets or triplets
of $SU(3)_C$,  i.e. we eschew quixes, queights, etc.,

4) all gauge anomalies vanish,

5) the standard model gauge groups are not embedded in the ETC group,

6) the ETC gauge group is asymptotically free,

7) the SSVZ mechanism is incorporated in order to keep the $\nu_\tau$
   light,

8) isospin and CP are not explicitly broken.

With these restrictions we can proceed straightforwardly.
Starting with $SU(2)_{TC}$, the simplest way to gauge the family
symmetries
is to make use of $SU(5)_{ETC}$. In order to get a hierarchy of
families, this gauge group should break down in stages (i.e.
$SU(5)_{ETC} \rightarrow SU(4)_{ETC} \rightarrow SU(3)_{ETC} \rightarrow
SU(2)_{TC}$).  In
order to avoid a visible techniaxion, as discussed
above, quarks and leptons are unified using the Pati-Salam \cite{PS}
group $SU(4)_{PS}$. In order to break $SU(4)_{ETC}$ and $SU(3)_{ETC}$
down to
$SU(2)_{TC}$, we will need an additional strong gauge group:
$SU(2)_{HC}$.  Thus, the gauge group
for the model is taken to be
$SU(5)_{ETC} \otimes SU(2)_{HC} \otimes SU(4)_{PS} \otimes SU(2)_{L}
\otimes U(1)_R$.   The breaking scale for all these interactions will be on the
order of
1000 TeV or lower.

To insure that the model contains only
${\bf 3}$'s, ${\bf \overline{3}}$'s, and singlets of color,
fermions are placed only in
antisymmetric irreducible representations \cite{Georgi} of
$SU(4)_{PS}$.
As usual, the $U(1)_R$
is required in order to get the correct hypercharges for the
right-handed
fermions.  Since, at the Pati-Salam breaking scale, $\Lambda_{PS}$,  the
$U(1)_R$
will mix with a generator of  $SU(4)_{PS}$ (with $\alpha_{PS}(\Lambda_{PS})
\approx 0.07$), the $U(1)_R$
coupling must be very weak in order to get the right $U(1)_Y$ coupling
in the low-energy effective theory.
The $U(1)_R$ gauge group looks like a remnant of an
$SU(2)_R$, but left-right symmetry has not been introduced
since we expect
that the requirement that the $SU(2)_L$ and $SU(2)_R$ gauge
couplings be
equal at the $SU(2)_R$ breaking scale would put this scale much
higher than those being considered here.   The reason for this is that the
$U(1)_R$ coupling at $\Lambda_{PS}$ is much weaker than the $SU(2)_L$ coupling
at this scale.

The standard model fermions and one family of technifermions can be
contained in the following representations
\footnote{Throughout we will make use of the convention of using the
charge-conjugates of the right-handed fields instead of the
right-handed fields themselves.}:
\begin{equation}
\begin{array}{lll}
({\bf 5},{\bf 1},{\bf 4},{\bf 2})_0 &
({\bf \overline{5}},{\bf 1},{\bf \overline{4}},{\bf 1})_{-1} &
({\bf \overline{5}},{\bf 1},{\bf \overline{4}},{\bf 1})_{1}  ~.
\end{array}
\end{equation}
If these were the only fermions in the model, there would be no
isospin splittings and no CKM mixing angles.  Thus we must include
additional fermions that can
mix with some of the ordinary fermions, so that
isospin breaking can arise spontaneously.

To motivate the choice of additional fermions, we next
consider how to include CP violation in the model, without
producing a strong-CP problem.  This can be done if
the Nelson-Barr solution to the strong-CP problem
\cite{CP} can be  implemented in our model.   The Nelson-Barr mechanism allows
complex phases to appear in the
mass mixing between the standard model quarks and new exotic quarks.  The
determinant
of the mass matrix, however, must remain real.   To begin, this mechanism
requires,
in addition to the standard fermions already discussed, some exotic
quarks that can mix with the ordinary quarks.  These quarks should
be $SU(2)_L$ singlets, so as not to contribute to $S$.  The simplest
way to do this (keeping in mind the restriction to antisymmetric
representations)
is to include  particles that transform as
$({\bf 6},{\bf 1})_0$ under $SU(4)_{PS} \otimes SU(2)_L \otimes
U(1)_R$.  Such representations
will decompose into particles with standard model quantum
numbers $({\bf \overline{3}},{\bf 1})_{2/3}$
and $({\bf 3},{\bf 1})_{-2/3}$. These correspond respectively
to a charge-conjugate, right-handed, down-type quark, and a left-handed
partner with which it can obtain a gauge invariant mass.  One such ``vector"
quark and
one hypercolored ``vector" quark will
be included, which we will refer to as the $m$ and the $G$ respectively.  The
$G$ will be responsible for feeding down a mass to the $m$, and will also slow
the running of the HC coupling above 10 TeV.  We will return to a  discussion
of CP violation in section 4.

We also need extra particles to incorporate the
SSVZ mechanism.  Since they must have the
quantum numbers of right-handed neutrinos, we can make use of the
simplest possibility: that they are $SU(4)_{PS} \otimes SU(2)_L \otimes
U(1)_R$ singlets. It can now be seen how isospin breaking can appear
spontaneously
in the  model.  The fermions of the standard model come from {\bf 4}'s and
${\bf \overline{4}}$'s of $SU(4)_{PS}$.  The additional Pati-Salam
representations
to be included are {\bf 1}'s and {\bf 6}'s, and these give
only right handed neutrinos and ``vector" down-type quarks.
Thus there will
be extra particles that can  mix with neutrinos and down-type quarks,
allowing for isospin breaking masses, and mixing angles.

We now explicitly write down the model.  The gauge group
is $SU(5)_{ETC}$ $\otimes SU(2)_{HC} \otimes
SU(4)_{PS}\otimes SU(2)_L \otimes U(1)_R$, with the
fermion content taken to be:
\begin{equation}
\begin{array}{lll}
({\bf 5},{\bf 1},{\bf 4},{\bf 2})_0
&
({\bf \overline{5}},{\bf 1},{\bf \overline{4}},{\bf 1})_{-1}
&
({\bf \overline{5}},{\bf 1},{\bf \overline{4}},{\bf 1})_{1}
\end{array}\nonumber
\label{ferm1}
\end{equation}
\begin{equation}
\begin{array}{ll}
({\bf1 },{\bf 1},{\bf 6},{\bf 1})_0 & ({\bf1 },{\bf 2},{\bf 6},{\bf 1})_0\\
\end{array}
\label{ferm2}
\end{equation}
\begin{equation}
\begin{array}{ll}
({\bf 10},{\bf 1},{\bf 1},{\bf 1})_0 &
 ({\bf 5},{\bf 1},{\bf 1},{\bf 1})_0 \\
({\bf \overline{10}},{\bf 2},{\bf 1},{\bf 1})_0 & ~.
\end{array}
\label{ferm3}
\end{equation}
The $({\bf 5},{\bf 1},{\bf 4},{\bf 2})_0$, the
$({\bf \overline{5}},{\bf 1},{\bf \overline{4}},{\bf 1})_{-1}$, and the
$({\bf \overline{5}},{\bf 1},{\bf \overline{4}},{\bf 1})_{-1}$ in this
list
contain particles with
quantum numbers corresponding to three families of ordinary
fermions
(plus charge-conjugated, right-handed neutrinos) and one family of
technifermions, i.e. the ${\bf 5}$ of $SU(5)_{ETC}$
corresponds to three families and two technicolors.
The additional fermions are an economical set that will allow us
to break ETC gauge symmetries, and isospin, as well as to incorporate the
Nelson-Barr mechanism for  CP violation.  Note
that the extra neutrino sector listed in (\ref{ferm3}) makes this a
chiral gauge theory with respect to the gauge groups $SU(5)_{ETC}$ and
$SU(2)_{HC}$.   All the non-Abelian gauge interactions  in the model are
asymptotically free.

Next the pattern of symmetry breaking must be specified.
An attractive and economical idea is that
the breaking is completely dynamical, driven by the asymptotically
free gauge theory itself at
each stage (this phenomena is referred to as ``tumbling" \cite{MAC1}). Folklore
then has it that the fermion
condensates form in the most attractive channel (MAC)
\cite{MAC1,MAC2}. The MAC is usually determined in one gauge boson
exchange
approximation, neglecting gauge boson masses that will be formed if
the
condensate breaks the gauge group.
The one-gauge-boson exchange approximation may or may not be reliable\footnote{
Some evidence for the reliability of the ladder
approximation is discussed in ref. \cite{ALM}.\label{ladderfoot}}, and
furthermore, the additional approximation
of neglecting gauge-boson mass generation could be
misleading.  We will nevertheless adopt the MAC criterion here as a guideline.

We will argue that the breaking will in fact take place in  the
phenomenologically desired breaking channel at
the lower ETC scales (approximately 100 TeV and below).  For this purpose we
will require that the $SU(5)_{ETC}$ and $SU(2)_{HC}$ couplings are relatively
strong in order to drive the tumbling.  (By contrast, the other gauge groups in
the model, which produce the weakly coupled interactions of the standard model,
will be too feeble to drive dynamical symmetry breaking.)  At ETC scales of
about 1000 TeV and above, the phenomenologically correct breaking channel will
not be the MAC,  and it will be necessary to assume that the breaking occurs in
the desired channel.  We take this as evidence that our model is complete below
1000 TeV, but perhaps not complete at  higher scales.

Thus, to begin, we assume
that the relatively strong $SU(5)_{ETC}$ gauge interactions and some additional
new
physics from higher scales
trigger the formation of a condensate at the scale $\Lambda_{PS}$ somewhat
above 1000 TeV
in the attractive channel $({\bf
\overline{5}},{\bf 1},{\bf \overline{4}},{\bf 1})_{-1} \times
({\bf 5},{\bf 1},{\bf 1},{\bf 1})_0
\rightarrow ({\bf 1},{\bf 1},{\bf  \overline{4}},{\bf 1})_{-1}$.
This breaks the $U(1)_R$ and Pati-Salam  symmetry, leading to the gauge group
$SU(5)_{ETC}
\otimes SU(2)_{HC} \otimes SU(3)_{C} \otimes
SU(2)_L \otimes U(1)_Y$ below $\Lambda_{PS}$.
Hypercharge, $Y$, (normalized by $Q = T_{3L} + Y/2$)  is given by
\begin{equation}
Y = Q_R +  \sqrt{{ 8 \over 3}}\,\, T_{15} ~,
\label{Y}
\end{equation}
where the $SU(4)_{PS}$ generator $T_{15} = \sqrt{{3\over 8}} \, {\rm
diag}({1\over 3},{1\over 3},{1\over 3},-1)$
is the $B-L$ generator, and $Q_R$ is the $U(1)_R$ charge.
Note that the $ ({\bf 1},{\bf 1},{\bf  \overline{4}},{\bf 1})_{-1}$ condensate
will
give a large mass to the right-handed neutrinos that were unified with
up-type quarks. This avoids the problem of quark masses feeding
down to neutrino masses through Pati-Salam interactions, discussed in Section
2.

The MAC \footnote{In our model this condensate would break the $SU(2)_{HC}$
group. Thus this channel may be disfavored  given that  $SU(2)_{HC}$ is
relatively strong, since the broken HC gauge bosons will give a large positive
contribution to the energy of the corresponding vacuum.\label{HC}}
 for $SU(5)_{ETC}$ is ${\bf 10} \times {\bf \overline{10}} \rightarrow {\bf
1}$.
The (massless) one-gauge-boson approximation gives a crude measure of the
strength of the interaction, and we will use this as a guideline throughout the
paper.  In this approximation the
interaction strength in this channel is proportional to the difference of
Casimirs, $\Delta C_2  = C_2({\bf 10}) + C_2({\bf \overline{10}}) -
C_2({\bf 1}) = 36/5$.
By contrast, the channel in which condensation is  assumed here  is the second
most attractive channel (with respect to $SU(5)_{ETC}$) with $\Delta C_2 =
24/5$.  As pointed out above, some additional new
physics at $\Lambda_{PS}$ and above may be necessary to produce the condensate
in this channel.

The fermion content of the model
below the Pati-Salam breaking scale, $\Lambda_{PS}$, (labeled by $SU(5)_{ETC}
\otimes SU(2)_{HC} \otimes SU(3)_{C} \otimes
SU(2)_L \otimes U(1)_Y$) is:
\begin{equation}
\begin{array}{ll}
({\bf 5},{\bf 1},{\bf 3},{\bf 2})_{1 / 3}&
({\bf 5},{\bf 1},{\bf 1},{\bf 2})_{-1} \\
({\bf \overline{5}},{\bf 1},{\bf \overline{3}},{\bf 1})_{-4/ 3}
&\\
({\bf \overline{5}},{\bf 1},{\bf \overline{3}},{\bf 1})_{2 / 3}&
({\bf \overline{5}},{\bf 1},{\bf 1},{\bf 1})_{2}
\end{array}\nonumber
\label{ferm51}
\end{equation}

\begin{equation}
\begin{array}{ll}
({\bf 1},{\bf 1},{\bf 3},{\bf 1})_{-2 / 3} &
({\bf 1},{\bf 1},{\bf \overline{3}},{\bf 1})_{2 / 3} \\
({\bf 1},{\bf 2},{\bf 3},{\bf 1})_{-2 / 3} &
({\bf 1},{\bf 2},{\bf \overline{3}},{\bf 1})_{2 / 3}
\end{array}
\label{ferm52}
\end{equation}

\begin{equation}
\begin{array}{ll}
({\bf 10},{\bf 1},{\bf 1},{\bf 1})_0 &
({\bf \overline{10}},{\bf 2},{\bf 1},{\bf 1})_0  ~.
\end{array}
\label{ferm53}
\end{equation}
 We have not listed the $({\bf
\overline{5}},{\bf 1},{\bf 1},{\bf 1})_{0}$ and the
$({\bf 5},{\bf 1},{\bf 1},{\bf 1})_0$ which have gotten a large
Dirac mass from the dynamical symmetry breaking.  We note that, except for
$U(1)_Y$, all the remaining gauge groups are asymptotically free.

Next we assume that, at
$\Lambda_5 \approx 1000$ TeV, a condensate forms in the attractive channel
$({\bf 10},{\bf 1},{\bf 1},{\bf 1})_0$ $\times$ $ ({\bf 10},{\bf 1},{\bf
1},{\bf 1})_0
\rightarrow
({\bf \overline{5}},{\bf 1},{\bf 1},{\bf 1})_0$. The $SU(5)_{ETC}$ MAC at this
scale would again be ${\bf 10} \times
{\bf \overline{10}}\rightarrow {\bf 1}$ with  $\Delta C_2 = 36/5$.  This
condensate, however,  would break $SU(2)_{HC}$, and might be disfavored as
pointed out in footnote \ref{HC}.  The assumed breaking channel, ${\bf 10}
\times {\bf 10} \rightarrow {\bf \overline{5}}$, is almost
as strong with  $\Delta C_2 = 24/5$, and it does not break $SU(2)_{HC}$.   Note
that the channel  $({\bf  \overline{10}},{\bf 2},{\bf 1},{\bf 1})_0$
$\times ({\bf  \overline{10}},{\bf 2},{\bf 1},{\bf 1})_0
\rightarrow
({\bf 5},{\bf 1},{\bf 1},{\bf 1})_0$ is  not a Lorentz scalar\footnote{It is
assumed here that gauge theories do not spontaneously break Lorentz
invariance.}, while
$({\bf  \overline{10}},{\bf 2},{\bf 1},{\bf 1})_0$
$\times ({\bf  \overline{10}},{\bf 2},{\bf 1},{\bf 1})_0
\rightarrow
({\bf 5},{\bf 3},{\bf 1},{\bf 1})_0$
is in a repulsive channel
with respect to the $SU(2)_{HC}$ interactions, and will be prevented
from forming if this gauge interaction is moderately strong.  (Thus the $({\bf
\overline{10}},{\bf 2},{\bf 1},{\bf 1})_0$ should not develop a Majorana mass.)
The condensate $({\bf
\overline{5}},{\bf 1},{\bf 1},{\bf 1})_0$ breaks the gauge symmetry to
$SU(4)_{ETC}
 \otimes SU(2)_{HC} \otimes SU(3)_C \otimes SU(2)_L \otimes U(1)_Y$, and the
first family breaks off at this
scale.   The fermion content below 1000 TeV is
(labeled according to $SU(4)_{ETC} \otimes SU(2)_{HC} \otimes SU(3)_C \otimes
SU(2)_L
\otimes U(1)_Y$):
\begin{equation}
\begin{array}{llll}
({\bf 1},{\bf 1},{\bf 3},{\bf 2})_{1/3} &
({\bf 4},{\bf 1},{\bf 3},{\bf 2})_{1/3} &
({\bf 1},{\bf 1},{\bf 1},{\bf 2})_{-1} &
({\bf 4},{\bf 1},{\bf 1},{\bf 2})_{-1}\\
(u,d)_L & & (\nu_e,e)_L& \\
\\
({\bf 1},{\bf 1},{\bf \overline{3}},{\bf 1})_{-4/3} &
({\bf \overline{4}},{\bf 1},{\bf \overline{3}},{\bf 1})_{-4/3}&
  \\
u_R^c & & &\\
\\
({\bf 1},{\bf 1},{\bf \overline{3}},{\bf 1})_{2/3} &
({\bf \overline{4}},{\bf 1},{\bf \overline{3}},{\bf 1})_{2/3} &
({\bf 1},{\bf 1},{\bf 1},{\bf 1})_{2} &
({\bf \overline{4}},{\bf 1},{\bf 1},{\bf 1})_{2} \\
d_R^c & & e_R^c &
\end{array}
\label{ferm41}
\end{equation}

\begin{equation}
\begin{array}{ll}
 ({\bf 1},{\bf 1},{\bf 3},{\bf 1})_{-2 / 3} &
 ({\bf 1},{\bf 1},{\bf \overline{3}},{\bf 1})_ {2 / 3}
\\ m_L  & m_R^c \\
\\
 ({\bf 1},{\bf 2},{\bf 3},{\bf 1})_{-2 / 3} &
 ({\bf 1},{\bf 2},{\bf \overline{3}},{\bf 1})_{2 / 3}\\
G_L & G_R^c
\end{array}
\label{ferm42}
\end{equation}

\begin{equation}
\begin{array}{ll}
({\bf 4},{\bf 1},{\bf 1},{\bf 1})_0 &
\\
({\bf \overline{4}},{\bf 2},{\bf 1},{\bf 1})_0&
({\bf 6},{\bf 2},{\bf 1},{\bf 1})_0  ~.
\end{array}
\label{ferm43}
\end{equation}
The names of standard model fermions have been written beneath the
corresponding group representations (where $u_R^c = (u_R)^c$).  We
have also labeled the
exotic, ``vector" $m$ quarks which should mix with the down-type quarks, and
the
hypercolored ``vector" $G$ quarks.
Note that there is no $\nu_{e R}^c$.  We also note that all
remaining non-Abelian gauge groups are again asymptotically free.

The next stage of breaking will be driven
by the $SU(4)_{ETC}$ and $SU(2)_{HC}$ interactions. It will be argued to occur
in the attractive
channel $({\bf \overline{4}},{\bf 2},{\bf 1},{\bf 1})_0\times
({\bf 6},{\bf 2},{\bf 1},{\bf 1})_0 \rightarrow
({\bf 4},{\bf 1},{\bf 1},{\bf 1})_0$ at a scale taken to be around $\Lambda_{4}
\approx
100$ TeV.
This breaks the gauge symmetry to $SU(3)_{ETC}\otimes SU(2)_{HC}
\otimes SU(3)_C \otimes
SU(2)_L \otimes U(1)_Y$, and the second family splits off at this
scale. This channel is not the MAC for $SU(4)_{ETC}$ alone: ${\bf 6} \times
{\bf 6} \rightarrow {\bf  1}$ ($\Delta C_2 = 5$) and ${\bf 4} \times
{\bf \overline{4}} \rightarrow {\bf 1}$ ($\Delta C_2 = 15/4$) are more
attractive. Nevertheless, both
${\bf \overline{4}}\times {\bf 6} \rightarrow {\bf 4}$ for $SU(4)_{ETC}$
($\Delta C_2 = 5/2$),
and
${\bf 2}\times{\bf 2} \rightarrow {\bf 1}$ for $SU(2)_{HC}$ ($\Delta C_2 =
3/2$) involve very attractive interactions (the latter is in fact the MAC for
$SU(2)_{HC}$).
We will next argue that the sum of these two interactions favors the chosen
channel over all others.

 It is not difficult to see that the most competitive other channel is the one
involving the
$SU(4)_{ETC}$ MAC:
$({\bf 6},{\bf 2},{\bf 1},{\bf 1})_0 \times ({\bf 6},{\bf 2},{\bf 1},{\bf 1})_0
\rightarrow
({\bf 1}, {\bf 3}, {\bf 1}, {\bf 1})_0$. To compare these two channels, we
compute for each the sum of the gauge couplings evaluated at $\Lambda_4$,
squared and
weighted by the difference of Casimirs in the various channels.  It is this
combination that will appear in an effective potential, or gap equation
analysis. For the channel involving the $SU(4)_{ETC}$ MAC, we have
\begin{equation}
\Delta C_2({\bf 6} \times {\bf 6} \rightarrow {\bf 1}) \, \alpha_4(\Lambda_4) +
\Delta C_2({\bf 2} \times {\bf 2} \rightarrow {\bf 3}) \, \alpha_2(\Lambda_4) =
5\, \alpha_4(\Lambda_4) - {1 \over 2}\, \alpha_2(\Lambda_4) ~,
\label{MAC}
\end{equation}
while for the desired channel we obtain
\begin{equation}
\Delta C_2({\bf {\overline 4}} \times {\bf 6} \rightarrow {\bf 4}) \,
\alpha_4(\Lambda_4) +
\Delta C_2({\bf 2} \times {\bf 2} \rightarrow {\bf 1}) \, \alpha_2(\Lambda_4) =
{5 \over 2} \,\alpha_4(\Lambda_4) + {3 \over 2}\, \alpha_2(\Lambda_4) ~.
\label{BIGMAC}
\end{equation}
Thus if  $\alpha_2(\Lambda_4) > {5 \over 4}\, \alpha_4(\Lambda_4)$
then (\ref{BIGMAC}) will be larger than (\ref{MAC}), and the desired channel
will be preferred over the other. We assume that this is the case.  Note that
as long as  $\alpha_2(\Lambda_4)  < {5 \over 3} \,\alpha_4(\Lambda_4)$, then it
is still the $SU(4)_{ETC}$ interactions that make the dominant contribution to
the dynamical symmetry breaking in the desired channel.  A simple gap equation
analysis (with constant couplings) indicates that
dynamical symmetry breaking will proceed when $\Delta C_2({\bf {\overline 4}}
\times {\bf 6} \rightarrow {\bf 4}) \, \alpha_4(\Lambda_4) +
\Delta C_2({\bf 2} \times {\bf 2} \rightarrow {\bf 1}) \, \alpha_2(\Lambda_4)$
reaches a critical
value of $2 \pi/3$.  More sophisticated analyses that include the effects of
running and gauge
boson masses generally find that $2 \pi/3$ is an underestimate of the critical
value.

It is instructive to compare our analysis with a conventional MAC analysis,
where one would compare the $SU(4)_{ETC}$ MAC   with the $SU(2)_{HC}$ MAC, i.e.
compare the first term in (\ref{MAC}) with the second term in (\ref{BIGMAC}).
Then one would find that as long as $\alpha_2(\Lambda_4)  < {10 \over 3}
\,\alpha_4(\Lambda_4)$, the $SU(4)_{ETC}$ interaction in channel (\ref{MAC})
would be dominant.   The ${\bf 6} \times {\bf 6} \rightarrow {\bf 1}$ channel
would be preferred for condensation, which, for the range of couplings
discussed above, would be the opposite conclusion to our more refined analysis.
 To summarize, we have suggested that when two (or more) relatively strong
gauge interactions are at play, the favored breaking channel will be determined
by the sum of the interactions. As in the present example, the favored channel
need not be the one involving the MAC of the strongest single interaction. We
refer to the favored channel in this case as the BIG MAC.  We  assume that the
coupling
constants are in the correct range for the BIG MAC to be preferred.

The fermion content below $\Lambda_{4}$ (labeled according to
$SU(3)_{ETC}\otimes SU(2)_{HC}
\otimes SU(3)_C \otimes
SU(2)_L \otimes U(1)_Y$) is:
\begin{equation}
\begin{array}{llll}
2({\bf 1},{\bf 1},{\bf 3},{\bf 2})_{1/3} &
({\bf 3},{\bf 1},{\bf 3},{\bf 2})_{1/3} &
2({\bf 1},{\bf 1},{\bf 1},{\bf 2})_{-1} &
({\bf 3},{\bf 1},{\bf 1},{\bf 2})_{-1}\\
(u,d)_L,(c,s)_L & &(\nu_e,e)_L,(\nu_\mu,\mu)_L & \\
\\
2({\bf 1},{\bf 1},{\bf \overline{3}},{\bf 1})_{-4/3} &
({\bf \overline{3}},{\bf 1},{\bf \overline{3}},{\bf 1})_{-4/3}&
  \\
u_R^c,c_R^c & & &\\
\\
2({\bf 1},{\bf 1},{\bf \overline{3}},{\bf 1})_{2/3} &
({\bf \overline{3}},{\bf 1},{\bf \overline{3}},{\bf 1})_{2/3} &
2({\bf 1},{\bf 1},{\bf 1},{\bf 1})_{2} &
({\bf \overline{3}},{\bf 1},{\bf 1},{\bf 1})_{2}\\
d_R^c,s_R^c & & e_R^c,\mu_R^c &
\end{array}
\label{ferm31}
\end{equation}
\newline
\begin{equation}
\begin{array}{ll}
 ({\bf 1},{\bf 1},{\bf 3},{\bf 1})_{-2 / 3} &
 ({\bf 1},{\bf 1},{\bf \overline{3}},{\bf 1})_{2 / 3}\\
m_L & m_R^c \\
\\
 ({\bf 1},{\bf 2},{\bf 3},{\bf 1})_{-2 / 3} &
 ({\bf 1},{\bf 2},{\bf \overline{3}},{\bf 1})_{2 / 3}\\
G_L & G_R^c
\end{array}
\label{ferm32}
\end{equation}
\newline
\begin{equation}
\begin{array}{ll}
({\bf 1},{\bf 1},{\bf 1},{\bf 1})_0 &
({\bf 3},{\bf 1},{\bf 1},{\bf 1})_0 \\
\nu_{\mu R}^c & \\
\\
({\bf 1},{\bf 2},{\bf 1},{\bf 1})_0 &
({\bf \overline{3}},{\bf 2},{\bf 1},{\bf 1})_0\\
X & F ~~~~~~~~~~~~~.
\end{array}
\label{ferm33}
\end{equation}
All the non-Abelian gauge groups at this stage are asymptotically free.  We
expect that the
gauge couplings $\alpha_2(\Lambda_4)$ and $\alpha_3(\Lambda_4)$ are in the
neighborhood of $0.5$.  For example, the values  $\alpha_2(\Lambda_4) \approx
0.61$ and $\alpha_3(\Lambda_4) \approx 0.47$ are
consistent with the BIG MAC analysis described above.

We note that the correct fermion content is now in place to employ the SSVZ
mechanism.
Consider the technifermions and third generation fermions (which transform
under $SU(3)_{ETC}$).  The charge-conjugated, right-handed $\nu_\tau$ and
technineutrino are in
a ${\bf 3}$ (see line (\ref{ferm33})) of $SU(3)_{ETC}$ as opposed to
charge-conjugated,
right-handed, quarks and charged leptons, which are in
${\bf \overline{3}}$'s.
Note  that the particle that  will turn out to be the
$\nu_{\mu R}^c$ has come out of the extra neutrino sector.
We also note that  the original fermion content
(in lines (\ref{ferm2}) and (\ref{ferm3})) above the Pati-Salam breaking scale
did not
suffer from Witten's anomaly \cite{Witten} for the  $SU(2)_{HC}$ gauge group.
This ensures the presence of
the particle we have labeled $X$, which did not appear in the original
SSVZ toy model \cite{SSVZ}.

The final stage of ETC breaking occurs when the $SU(2)_{HC}$ and $SU(3)_{ETC}$
interactions get somewhat stronger, at a scale $\Lambda_{3}$ that will be
roughly estimated to be around $10$ TeV.
The desired channel is the one in which the $F$
condenses with itself: $({\bf \overline{3}}, {\bf {2}}, {\bf {1}}, {\bf {1}})_0
\times
({\bf \overline{3}}, {\bf {2}}, {\bf {1}}, {\bf {1}})_0 \rightarrow ({\bf {3}},
{\bf {1}}, {\bf {1}}, {\bf {1}})_0$, breaking $SU(3)_{ETC}$ to $SU(2)_{TC}$.
This is the MAC for $SU(2)_{HC}$, and an attractive channel for $SU(3)_{ETC}$.
The combination of the two interactions ensures that the $F$
condenses with itself rather than with the $X$ or the $G$, and provides another
example of a BIG MAC.  Again, as a guideline, we consider the sum of the gauge
couplings, squared and
weighted by the difference of Casimirs for this channel:
\begin{equation}
\Delta C_2({\bf \overline{3}} \times
{\bf \overline{3}} \rightarrow {\bf 3}) \, \alpha_3(\Lambda_3) +
\Delta C_2({\bf 2} \times {\bf 2} \rightarrow {\bf 1}) \, \alpha_2(\Lambda_3) =
{4 \over 3} \,\alpha_3(\Lambda_3) + {3 \over 2}\, \alpha_2(\Lambda_3) .
\label{BIGMAC23}
\end{equation}
Condensation should occur  when expression (\ref{BIGMAC23}) is about $2 \pi/3$.

Note that since the coefficient of $\alpha_3(\Lambda_3)$ in (\ref{BIGMAC23}) is
less than the coefficient of $\alpha_4(\Lambda_4)$ in (\ref{BIGMAC}),
$\alpha_3(\Lambda_3)$ must be larger than  $\alpha_4(\Lambda_4)$ in order for
dynamical symmetry breaking to occur at both  $\Lambda_3$ and $\Lambda_4$.
This is consistent with the asymptotic freedom of the $SU(3)_{ETC}$ and
$SU(2)_{HC}$ gauge groups in our model.  Some walking (recall that the
``vector"  $G$ quarks help to reduce the one-loop HC $\beta$-function) of the
HC gauge coupling will be required to make this condensation occur at a low
enough scale ($\approx  10$ TeV).

 For comparison, the MAC for
$SU(3)_{ETC}$ is ${\bf 3} \times {\bf \overline{3}} \rightarrow {\bf 1}$.
(Note that all the ${\bf 3}$'s of $SU(3)_{ETC}$ in the model are $SU(2)_{HC}$
singlets, so there is no possibility of additional interactions to assist the
condensation in this channel.)  For this channel, the
squared coupling weighted by the difference of Casimirs is:
\begin{equation}
\Delta C_2({\bf 3} \times {\bf \overline{3}} \rightarrow {\bf 1}) \,
\alpha_3(\Lambda_3) =
{8 \over 3} \, \alpha_3(\Lambda_3) ~.
\label{MAC3}
\end{equation}
Thus for $\alpha_2(\Lambda_3) > {8 \over 9} \alpha_3(\Lambda_3)$, expression
(\ref{BIGMAC23})
is larger than (\ref{MAC3}), and the breaking proceeds as required:  $({\bf
\overline{3}}, {\bf {2}}, {\bf {1}}, {\bf {1}})_0 \times
({\bf \overline{3}}, {\bf {2}}, {\bf {1}}, {\bf {1}})_0 \rightarrow ({\bf {3}},
{\bf {1}}, {\bf {1}}, {\bf {1}})_0$.

The condensate $({\bf {3}},
{\bf {1}}, {\bf {1}}, {\bf {1}})_0$ breaks the ETC gauge symmetry down  to TC:
$SU(2)_{TC}  \otimes SU(2)_{HC} \otimes SU(3)_C \otimes SU(2)_L \otimes
U(1)_Y$.  The component of $F$ that
is neutral under TC
does not get a mass from this condensate.  This component does however condense
with the $X$, at a slightly lower scale, $\Lambda_{HC}$.  The $G$ quarks will
also condense at
$\Lambda_{HC}$.  Since the HC coupling is quite strong at $\Lambda_3 \approx
10$  TeV, with
a standard running of this coupling  $\Lambda_{HC}$ will be very close to
$\Lambda_3$.
Hypercolored particles are confined at $\Lambda_{HC}$, and the HC sector
decouples from ordinary fermions and
technifermions.  We then have a one-family TC model, with an additional
``vector" quark, $m$.

The fermion content below $\Lambda_3 \approx 10$ TeV
(labeled according to $SU(2)_{TC} \otimes SU(3)_C$ $\otimes
SU(2)_L \otimes U(1)_Y$) is:
\begin{equation}
\begin{array}{llll}
3({\bf 1},{\bf 3},{\bf 2})_{1/3} &
({\bf 2},{\bf 3},{\bf 2})_{1/3} &
3({\bf 1},{\bf 1},{\bf 2})_{-1} &
({\bf 2},{\bf 1},{\bf 2})_{-1}\\

(u,d)_L,(c,s)_L,(t,b)_L & (U,D)_L &(\nu_e,e)_L,(\nu_\mu,\mu)_L,
(\nu_\tau,\tau)_L&(N,E)_L\\
\\
3({\bf 1},{\bf \overline{3}},{\bf 1})_{-4/3} &
({\bf 2},{\bf \overline{3}},{\bf 1})_{-4/3}&
  \\
u_R^c,c_R^c,t_R^c & U_R^c& &\\
\\
3({\bf 1},{\bf \overline{3}},{\bf 1})_{2/3} &
({\bf 2},{\bf \overline{3}},{\bf 1})_{2/3} &
3({\bf 1},{\bf 1},{\bf 1})_{2} &
({\bf 2},{\bf 1},{\bf 1})_{2} \\
d_R^c,s_R^c,b_R^c &D_R^c &e_R^c,\mu_R^c,\tau_R^c &E_R^c
\end{array}
\label{ferm21}
\end{equation}

\begin{equation}
\begin{array}{ll}
 ({\bf 1},{\bf 3},{\bf 1})_{-2 / 3} &
 ({\bf 1},{\bf \overline{3}},{\bf 1})_{2 / 3}\\
m_L & m_R^c
\end{array}
\label{ferm22}
\end{equation}

\begin{equation}
\begin{array}{ll}
2({\bf 1},{\bf 1},{\bf 1})_0 &
({\bf 2},{\bf 1},{\bf 1})_0\\
\nu_{\mu R}^c,\nu_{\tau R}^c & N_R^c ~~~~~~~~.
\end{array}
\label{ferm23}
\end{equation}
Note that ${\bf 2}$'s and ${\bf \overline{2}}$'s of $SU(2)_{TC}$
are equivalent.
All interactions except $U(1)_Y$ are asymptotically free.  The model at this
stage consists of the usual three families (left and right-handed except that
there is
no right-handed $\nu_{e}$), one conventional family of technifermions, and the
``vector" quark $m$.

At the technicolor scale $\Lambda_{TC}$, the $SU(2)_{TC}$ coupling becomes
strong enough that ${3 \over {2 \pi}}  \Delta C_2({\bf 2} \times {\bf 2}
\rightarrow {\bf 1})\, \alpha_{TC}(\Lambda_{TC})={\cal O}(1)$.  Technifermions
then get dynamical masses, $SU(2)_{L} \times
U(1)_{Y}$ breaks to $U(1)_{em}$, and the masses of the quarks and leptons are
generated by the ETC interactions linking the various particles in (\ref
{ferm21}) and (\ref{ferm23}).
We turn next to a description of these masses and other
features of the model.

\section{Features and Problems of the Model}

In this section we will discuss the mass spectrum of  ordinary fermions, some
of the phenomenology of the TC sector, and CP violation.  To begin we note
that the three, well-separated,
ETC scales  in the model provide a natural starting point for an explanation of
the pattern of
family masses.  Furthermore, as
we will discuss in more detail below, it is possible that  QCD interactions
will adequately split quark masses from lepton masses.  The combination of the
above mentioned effects with  the SSVZ mechanism (which suppresses
neutrino masses) will then generate  the overall gross features of the quark
and lepton spectrum.

We first discuss the masses of the third family ($t$,$b$,$\nu_{\tau}$,$\tau$).
 We note  that in a moderately walking TC theory, $\Lambda_3 \approx 10$ TeV is
a natural scale to generate the mass of the $\tau$.  In order
to explain the $t-\tau$ hierarchy of nearly two orders of magnitude it will be
assumed
that the ETC interactions linking the $t$ to the $U$ are near-critical
\cite{strongETC} at $\Lambda_{3}$, i.e.  $ \alpha_3(\Lambda_3)$ is very close
to, but below, a critical value $\alpha_c(3)$, given by a
crude Schwinger-Dyson equation (in the ladder approximation\footnote{See
footnote \ref{ladderfoot}.}) analysis to be
\begin{equation}
\alpha_c(3)={{2 \pi} \over 3 \,\,  \Delta C_2({\bf 3} \times {\bf \overline{3}}
\rightarrow {\bf 1}) }
= {{\pi}\over{4}} ~.
\label{nearcrit}
\end{equation}
Then as pointed out in Section 2, the additional effect of the QCD interaction
in the gap equation for the $t$ and $U$ can dramatically enhance $m_t$ relative
to $m_\tau$ \cite{color2}.
In particular, if
$\alpha_3(\Lambda_3)$ is within $1$-$10\%$ of $\alpha_c(3)$, then it is
possible to produce an $m_t$ in the $150$ GeV range with $m_\tau = 1.8$ GeV.
We note that if $\alpha_3(\Lambda_3)$ is near-critical, then the mass of the
techniquarks (which sets the scale for the $W$ and $Z$ masses) can be
substantially larger than the intrinsic TC scale, $\Lambda_{TC}$.  The scale
$\Lambda_{TC}$ could be as low as 100 GeV \cite{revenge}.  Since
$\alpha_{TC}(\Lambda_3) = \alpha_3(\Lambda_3)$, the TC coupling must be
moderately walking from
 $\Lambda_3$ down to $\Lambda_{TC}$  in order for $\Lambda_{TC}$ to be much
smaller than $\Lambda_3$.  For the TC coupling in this range,  the perturbative
expansion for the $\beta$ function may be unreliable.  The same is true for
some of the ETC and HC gauge couplings relevant at higher scales. In this
paper, we will not attempt to compute these $\beta$ functions. Instead, we will
simply point out the qualitative behavior that is necessary in each energy
range.

Consider the ETC and HC couplings in the range from  $\Lambda_4 \approx 100$
TeV to $\Lambda_3 \approx 10$ TeV.  Suppose, as discussed in section 3, that
the $SU(2)_{HC}$ coupling $\alpha_2(\Lambda_4)=0.61$, and
$\alpha_4(\Lambda_4)=0.47$.  This makes expression (\ref{BIGMAC}) equal to $2
\pi/3$, and $\alpha_2(\Lambda_4) > {5 \over 4} \, \alpha_4(\Lambda_4)$.   For
the model to work, the coupling $\alpha_3$ must walk from $\Lambda_4$  to
$\Lambda_3$ be near $\alpha_c(3) = \pi/4 \approx 0.8$ at $\Lambda_3$.   Also
$\alpha_2$ must  be walking in order for $\Lambda_3$ to be an order of
magnitude smaller than $\Lambda_4$.

In order to estimate masses of the quarks and leptons, we need  estimates for
the condensates of the technifermions.  However our model is far from QCD-like,
so we cannot simply scale-up the QCD condensate.   Instead we use the $t$ and
$\tau$ masses as inputs to determine the relevant condensates, and use these
estimates to calculate the masses of  particles in the second and first
families.
We  expect that the $\tau$ mass is given roughly by the standard
one-ETC-gauge-boson-exchange graph:
\begin{equation}
m_\tau \approx  3 \pi \alpha_3(\Lambda_3) \,
{{<  {\overline E_R} E_L> }\over{\Lambda_3^2}}
 ~.
\label{mtau}
\end{equation}
The coefficient  $3 \pi \alpha_3(\Lambda_3)$ can be understood as follows.  The
one-ETC-gauge-boson-exchange graph is given by $3  \alpha_3(\Lambda_3)  C/(4
\pi)$ times
an integral of the technielectron self-energy.  This integral is $4\pi^2$ times
the technielectron condensate\footnote{Our convention for the condensate is the
negative of the more usual convention.} $<  {\overline E_R} E_L>$ .  The
constant $C$ comes from the squares of ETC generators, and for the
representations in our model turns out to be $N/2$, where $N$ is the number of
heavy ETC gauge bosons which contribute to the graph.  For $SU(3)_{ETC}
\rightarrow SU(2)_{TC}$, $N = 2$; for $SU(4)_{ETC} \rightarrow SU(3)_{ETC}$, $N
= 3$, and so on.  Thus, rewriting equation (\ref{mtau}), we take the
charged-technilepton condensate to be:
\begin{eqnarray}
<  {\overline E_R} E_L>  &\approx&  {4 \over 3 \pi^2} \,\,m_\tau \Lambda_3^2
\nonumber\\
&\approx& 0.024 \,{\rm TeV}^3
 ~.
\label{EE}
\end{eqnarray}

Since the mass of the $t$ is comparable to the techniquark mass, the
corresponding Schwinger-Dyson equations are  near-critical and non-linear, and
we do not expect a simple formula like (\ref{mtau}) to apply for $m_t$.   We
expect that just below $\Lambda_3$, the dynamical mass of the $U$ techniquark,
$\Sigma_U$,  is roughly constant (for a larger range than the technilepton
mass), and is close to $m_t$.  Thus, we will simply use the estimate:
\begin{eqnarray}
<  {\overline U_R} U_L> &\approx& {{1}\over{4 \pi^2}}
\int_0^{\Lambda_3^2}\,dk^2\,
{{k^2\, \Sigma_U(k)}\over{k^2 +\Sigma^2_U(k)}} \nonumber\\
&\approx& {{1}\over{4 \pi^2}} \int_0^{\Lambda_3^2}\,dk^2\, m_t  \nonumber\\
&\approx& {{m_t \Lambda_3^2}\over{4 \pi^2}} \nonumber\\
&\approx& 0.38\,{\rm TeV}^3
 ~,
\label{UU}
\end{eqnarray}
where we have made the approximation that the integral is dominated at momenta
near $\Lambda_3$, and taken $m_t = 150$ GeV.

The mass of the $\nu_\tau$ is suppressed as in the SSVZ mechanism
described in Section 2.  While the $E_L$ and $E_R^c$  transform
as a  ${\bf 3}$
and a ${\bf \overline{3}}$ under $SU(3)_{ETC}$, the $N_L$ and
$N_{R}^c$ both  transform as ${\bf 3}$'s.
Thus, a Dirac mass will not feed down to the
$\nu_\tau$ unless there is some mixing of
ETC gauge bosons, since the one ETC gauge boson exchange graph is
identically zero.  The $\nu_\tau$ does not receive a mass even at two loops.
In fact one can show that a mass cannot feed down
to the $\nu_\tau$ from the technineutrino mass alone, to all orders in
perturbation theory.  The reason for this is that the technineutrino
mass transforms as part of a ${\bf 3}$ of $SU(3)_{ETC}$, while the
$\nu_\tau$
mass transforms as part of a ${\bf 6}$; the appropriate component of the
${\bf 6}$ can only be made from an even number of ${\bf 3}$'s, but there must
be an odd number of mass insertions in order to have a helicity
flip.    We expect, however,  that at three loops particles other
than neutrinos can feed down a mass to the $\nu_\tau$.  As we will see however,
there are more
important effects that will couple the $\nu_{\tau R}^c$ to the $\nu_e$.  We
will return to this  when we discuss the first generation.

The remaining member of the third family is the $b$ quark.
The mechanism for generating its mass is quite different from that for the $t$
quark. The $t$ gets its mass only through the standard
one ETC gauge boson exchange, while the $b$ mass can be suppressed by
mixing with the $m$ quark.  Since the Schwinger-Dyson equations for the mass of
the $b$ and the mass of the $D$ techniquark are coupled, the reduced $b$ mass
feeds back into the mass
(renormalized near the ETC scale, $\Lambda_3 \approx 10$ TeV) of the
$D$ techniquark, which lowers its mass, and further lowers
the $b$ mass. Thus
this model may not have a problem accommodating a large $t$-$b$ mass
splitting.   The calculation of the $b$ quark mass will require further
information about the mixing with the $m$ quark, which depends on physics at
and above 1000 TeV.

With the interactions discussed so far, the $m$ quark remains massless.  In
order for it to gain a
mass, and to mix with the down-type quarks, there must be additional physics,
which will take
the form of higher dimension operators in the low-energy effective theory below
$\Lambda_5 \approx 1000$ TeV. An example of an operator that would give the $m$
a mass is:
\begin{equation}
{\cal L}_{4f} = { g^2 \over {\Lambda_5^2}} \,{\overline G_R} G_L {\overline
m_R} m_L +
{\rm hermitian \,\,conjugate} ~,
\label{4f}
\end{equation}
where we expect $g^2/4 \pi$ to be ${\cal O}(1)$.
Then when the $G$ gets a mass at $\Lambda_{HC}$,  this mass will feed down to
the $m$
through the four-fermion operator (\ref{4f}).  In order to estimate this mass,
we will need the value of the condensate $< {\overline G_R} G_L>$ cut off at
the scale $\Lambda_5$.  We recall   that the anomalous dimension of  the mass
operator (in ladder approximation\footnote{See footnote \ref{ladderfoot}.}) in
an $SU(N)$ gauge theory is:
\begin{equation}
\gamma_N(\alpha) = 1 - \sqrt{1 - {{\alpha}\over{\alpha_c(N)}}} ~,
\label{gamma}
\end{equation}
where $\alpha_c(N)$ is the generalization of equation (\ref{nearcrit}) to the
appropriate gauge group, and we are assuming  $\alpha<\alpha_c(N)$.  We also
recall that for an extremely slowly running coupling between the symmetry
breaking  scale $\mu$ and a larger scale $\Lambda$, the condensate $<{\overline
\psi} \psi >$ cut off at $\Lambda$ is roughly given by
\begin{equation}
<{\overline \psi} \psi >_\Lambda \, \,\approx\,  <{\overline \psi} \psi >_\mu
\left( {{\Lambda}\over{\mu}}\right)^{\gamma_N(\alpha)} ~.
\label{enhance}
\end{equation}

Of course the coupling does run;  for the purposes of a crude calculation we
will use an average coupling ${\overline\alpha}$.
In order to make an estimate, we  split the range of momenta into two, from
$\Lambda_{HC} \approx \Lambda_3 \approx 10$ TeV to  $\Lambda_4 \approx 100$
TeV, and from $\Lambda_4$ to $\Lambda_5 \approx 1000$ TeV.   We expect
$\alpha_2$ to run from $\alpha_2(\Lambda_{HC})=\alpha_{c}(2) = 4 \pi/9 \approx
1.4$,  to
$\alpha_2(\Lambda_4)\approx 0.6$ (as discussed above) over the lower range.  We
will take $\alpha_2(\Lambda_5)= 0.4$.  Thus we have ${\overline\alpha_2}=1$
over the lower range, and ${\overline\alpha_2}=0.5$ over the upper range.  We
assume that  the  $< {\overline G_R} G_L>$ condensate is at least as big as a
scaled-up QCD condensate (i.e. $4 \pi f^3$).  The mass of the $m$ is then:
\begin{eqnarray}
m_m &\approx&  { g^2 \over {\Lambda_5^2}}  < {\overline G_R} G_L>
\left( {{\Lambda_4}\over{\Lambda_3}}\right)^{\gamma_2({\overline\alpha_2}=1.0)}
\left( {{\Lambda_5}\over{\Lambda_4}}\right)^{\gamma_2({\overline\alpha_2}=0.5)}
\nonumber \\
 & \approx & { g^2 \over {\left(1000 \,{\rm TeV} \right)^2}}  \,\, 4 \pi
\left(10\, {\rm TeV}\right)^3
\left( {{100\, {\rm TeV}}\over{10 \,{\rm TeV}}}\right)^{0.5}
\left( {{1000\, {\rm TeV}}\over{100 \,{\rm TeV}}}\right)^{0.2} \nonumber\\
& \approx  &  g^2 \, 60\, {\rm GeV} ~.
\label{mm}
\end{eqnarray}
With $g^2/4\pi > 0.2$, our estimate for $m_m$ is above the current experimental
lower bound  ($\approx 110$ GeV) for such a particle.  Of course $m_m$ does not
correspond to the physical mass of the $m$, since it must mix with the
down-type quarks, and this could change the value of the physical mass.

It is also important to comment on the masses of the single  family of
technifermions in this model.
With the ETC coupling at $\Lambda_3$ close enough to criticality so that the
$t$ is much heavier than
the $\tau$, the techniquarks will be much heavier than technileptons,
\cite{color2}.
Also, since the technielectron has attractive ETC interactions in the scalar
channel while
the technineutrino has repulsive ETC interactions, the technielectron
will be heavier than the technineutrino.  Thus this model can provide
a realization of the technifermion mass pattern suggested in
ref. \cite{revenge}.  It was shown there that with this breaking of $SU(2)_R$
the electroweak radiative  correction parameter $S$ will be
smaller than is estimated in QCD-like TC models, and may even be
negative.  We also expect that the lightness of the technineutrinos will
lead to a very light techni-$\rho$ (composed of  technineutrinos and
antitechnineutrinos), that may be light enough to be seen at LEP II
\cite{revenge}.  This model will also generate a significant ($m_t$ dependent)
correction \cite{Zbb} to the $Z \rightarrow b {\overline b}$ vertex, which
should be accurately measured soon.
The spectrum of pseudo-Nambu-Goldstone bosons should also be similar\footnote{
However there will be two more pseudo-Nambu-Goldstone bosons in the model
discussed here, since there is no distinction
between $N_L$ and $N_R^c$.  We leave a detailed examination of the pseudos for
future work.} to that
sketched out in ref. \cite{revenge}.

Later in this section we will need an estimate of the technineutrino
condensate, in order to estimate ordinary neutrino masses.  Since the
technineutrinos have repulsive
ETC interactions, the integral representing the technineutrino
condensate should converge rapidly above 100 GeV (which we take as an order
of magnitude
estimate of the technineutrino mass \cite{revenge}).
So we take
\begin{eqnarray}
 <  {\overline N_R} N_L>&\approx& {{1}\over{4 \pi^2}}
\int_0^{\Lambda_3^2}\,dk^2\,
{{k^2 \, \Sigma_N(k)}\over{k^2 + \Sigma^2_N(k)}} \nonumber\\
&\approx&  {{(100 {\rm GeV})^3}\over {8 \pi^2}} \nonumber\\
&\approx&1.3 \times 10^{-5}\, {\rm TeV}^3~.
\label{NN}
\end{eqnarray}

We also note that the vacuum alignment
problem \cite{vac} of $SU(2)_{TC}$ theories, with one family of degenerate
technifermions,
should not be present in this model.  Recall that,
in the absence of ETC interactions, the contribution of the (unbroken)
electroweak
gauge bosons to the vacuum energy causes the $N_L$ to condense  with the $E_L$
rather
than $N_R^c$.  This technilepton condensate breaks
$U(1)_{em}$ rather than $SU(2)_L$, which obviously does not correspond to the
observed
vacuum\footnote{By contrast, the techniquark condensate breaks electroweak
gauge symmetry in the correct fashion due to the presence  of QCD interactions
\cite{vac}.}.  In our model, however,
the strong ETC interactions will lower the energy of the vacuum where $E_L$
condenses with
$E_R^c$ (cf. ref. \cite{Sundrum}).  Moreover since the techniquarks condense at
a higher energy scale, at the scale where the technileptons condense, the
$W^\pm$ and $Z$ have already
gotten the bulk of their masses from the techniquark condensate, and hence
their (destabilizing) contribution to the vacuum energy will be suppressed.

We turn next to a discussion of the second family ($c$,$s$,$\nu_\mu$,$\mu$).
With a moderate enhancement from walking, the mass of the $\mu$ can be obtained
naturally with an ETC scale of $\Lambda_4 \approx 100$ TeV.  We know that
$\alpha_3$ must run from $\alpha_3(\Lambda_3) \approx \alpha_c(3) \approx 0.79$
to $\alpha_3(\Lambda_4)= \alpha_4(\Lambda_4) \approx 0.47$ (as discussed
earlier), so we take ${\overline \alpha_3}=0.7$.  We then have:
\begin{eqnarray}
m_\mu &\approx&  { {9  \pi \alpha_4(\Lambda_4)} \over { 2 }}  \,
{{<  {\overline E_R} E_L> }\over{\Lambda_4^2}}
\left( {{\Lambda_4}\over{\Lambda_3}}\right)^{\gamma_3({\overline \alpha_3})}
\nonumber \\
 & \approx &6.7  \,\,{ { 0.024 {\rm TeV}^3} \over {\left(100\, {\rm TeV}
\right)^2}}
\left( {{100\, {\rm TeV}}\over{10 \,{\rm TeV}}}\right)^{0.67}
\nonumber\\
& \approx  &  100\, {\rm MeV} ~.
\label{mmu}
\end{eqnarray}

The same physics that gives a large $t$ mass will also enhance the $c$ mass
relative to the $\mu$.  Assuming that the correct $t$ mass is generated, as
discussed above, we can roughly estimate the $c$ mass as:
\begin{eqnarray}
m_c &\approx&  { {9 \pi \alpha_4(\Lambda_4)} \over {2 }}
\, {{<  {\overline U_R} U_L> }\over{\Lambda_4^2}}
\left( {{\Lambda_4}\over{\Lambda_3}}\right)^{\gamma_3({\overline \alpha_3})}
\nonumber \\
 & \approx &6.7\,\,  { {0.38\, {\rm TeV}^3} \over {\left(100\, {\rm TeV}
\right)^2}}
\left( {{100\, {\rm TeV}}\over{10\,  \rm{TeV}}}\right)^{0.67}
\nonumber\\
& \approx  &  1 \,{\rm GeV} ~.
\label{mc}
\end{eqnarray}
The results for $m_\mu$ and $m_c$ are quite good for such crude estimates.  One
could hope to do better with a more refined analysis of the Schwinger-Dyson
equations.
We further expect that mixing
with the $m$ quark will reduce the mass of the $s$ quark, just as in the case
of the $b$ quark.

The $\nu_\mu$ is the heaviest neutrino in our model.  It does
not receive a mass at one loop, but it does at two  loops.  The
extra loop is necessary to mix two different 100 TeV ETC gauge
bosons.  The mixing breaks $SU(3)_{ETC}$ and $SU(4)_{ETC}$, and so
should be of order $10\, {\rm TeV} \times 100\,{\rm TeV}$.  Thus we expect the
$\nu_\mu$
neutrino mass to be given roughly by:
\begin{eqnarray}
m_{\nu_\mu} & \approx &18 \pi^2 \alpha_4^2(\Lambda_4)\,
{{ <  {\overline N_R} N_L>}\over{\Lambda_4^4}}
{{\Lambda_3 \Lambda_4}\over {16\pi^2}} \nonumber \\
& \approx & 40 \,\, {{1.3\times10^{-5}\, {\rm TeV}^3}\over{(100\, {\rm
TeV})^3}}
{{10\, {\rm TeV}}\over {16\pi^2}}\nonumber\\
& \approx &  30\, {\rm eV} ~.
\label{mnumu}
\end{eqnarray}
Note that the coefficient in equation (\ref{mnumu}) is $g_4^2(\Lambda_4) = 4
\pi \alpha_4(\Lambda_4)$ times that in equations (\ref{mmu}) and (\ref{mc}),
since there is an extra ETC gauge boson exchange.  The $1/16 \pi^2$ is the
standard estimate of the suppression due to an extra loop.
It is interesting that $30$ eV is the right mass for a stable Dirac neutrino to
close the universe, but considerations of structure formation indicate that a
lighter neutrino mass is preferred.    However, the neutrino mass estimates in
our model are more unreliable than those of other fermions, since the neutrino
masses only arise at two loops, and there is, as yet, no experimental input to
determine the technineutrino condensate in equation (\ref{NN}).

Finally we briefly discuss the first family ($u$,$d$,$\nu_e$,$e$).  An ETC
scale of roughly $\Lambda_5 \approx 1000$ TeV will be sufficient to give
naturally the correct mass for the $e$.
To see this, we again split the range of momenta into two parts, from
$\Lambda_3$ to  $\Lambda_4$, and from $\Lambda_4$ to $\Lambda_5$.    As
discussed above, $\alpha_4(\Lambda_4)= 0.47$, and we take $\alpha_4(\Lambda_5)
=\alpha_5(\Lambda_5) = 0.1$.
Thus we have $ {\overline \alpha_4}=0.35$.  A crude calculation then gives
 \begin{eqnarray}
m_e &\approx&  6 \pi \alpha_5(\Lambda_5)\,
{{< {\overline E_R} E_L>}\over{ \Lambda_5^2}}
\left( {{\Lambda_4}\over{\Lambda_3}}\right)^{\gamma_3({\overline \alpha_3})}
\left( {{\Lambda_5}\over{\Lambda_4}}\right)^{\gamma_4({\overline \alpha_4})}
\nonumber \\
 & \approx & 1.9\,\,  { {0.024\,{\rm TeV}^3}\over {\left(1000 \,{\rm TeV}
\right)^2}}
\left( {{100\, {\rm TeV}}\over{10\, {\rm TeV}}}\right)^{0.67}
\left( {{1000 \,{\rm TeV}}\over{100\, {\rm TeV}}}\right)^{0.32} \nonumber\\
& \approx  &  1\, {\rm MeV} ~.
\label{me}
\end{eqnarray}
Note that most of the walking enhancement comes from the momentum range 10 TeV
to 100 TeV.

The QCD
enhancement of the $t$ and $c$ quark masses, discussed above, will put the $u$
and $d$ masses in the right range of $5 - 10$ MeV:
\begin{eqnarray}
m_u &\approx&  6 \pi  \alpha_5(\Lambda_5)
\, {{<  {\overline U_R} U_L> }\over{\Lambda_4^2}}
\left( {{\Lambda_4}\over{\Lambda_3}}\right)^{\gamma_3({\overline \alpha_3})}
\left( {{\Lambda_5}\over{\Lambda_4}}\right)^{\gamma_4({\overline \alpha_4})}
\nonumber \\
 & \approx &1.9\,\, { {0.38 \,{\rm TeV}^3} \over {\left(1000\, {\rm TeV}
\right)^2}}
\left( {{100 \,{\rm TeV}}\over{10 \, \rm{TeV}}}\right)^{0.67}
\left( {{1000 \,{\rm TeV}}\over{100 \,{\rm TeV}}}\right)^{0.32} \nonumber\\
& \approx  &  10 \,{\rm MeV} ~.
\label{mu}
\end{eqnarray}
The estimates for $m_e$ and $m_u$ are encouraging, and again suggest that a
more refined analysis of the Schwinger-Dyson equations is merited.
The size and sign of the $u-d$ mass splitting remains unexplained so far.  It
must arise from mixing with the $m$ quark driven by additional, high energy
interactions.

At two loops,  $\nu_{e L}$ gets a mass with the $\nu_{\tau R}^c$.  As with the
$\nu_\mu$ we must mix two different ETC gauge bosons, but in this case only one
is associated with  $SU(5)_{ETC}$ breaking (and thus has a mass around
$\Lambda_5$), while the other is associated with $SU(3)_{ETC}$ breaking (and
thus has a mass around $\Lambda_3$).  The mixing term requires
two $SU(3)_{ETC}$ breaking dynamical masses, one from the $X-F$ mass, and one
from the $F$ mass.  Thus (taking $\alpha_3(\Lambda_3)=0.79$, and
$\alpha_5(\Lambda_5)=0.1$) we have
\begin{eqnarray}
m_{\nu_e} & \approx &12 \pi^2 \alpha_3(\Lambda_3)\alpha_5(\Lambda_5)\,
 {{ <  {\overline N_R} N_L>}\over{\Lambda_5^2\,\Lambda_3^2}}
{{\Lambda_3^2 }\over {16\pi^2}} \nonumber \\
& \approx & 9.4 \, {{1.3 \times 10^{-5}\, {\rm TeV}^3}\over{(1000\, {\rm
TeV})^2}}
{1 \over {16\pi^2}}\nonumber\\
&\approx & 1\, {\rm eV} ~.
\label{mnue}
\end{eqnarray}
Thus the $\nu_{\tau R}^c$ becomes part of a Dirac neutrino, the $\nu_e$.

To recap the neutrino sector, at two-loop order we have two Dirac neutrinos
($\nu_\mu$ and $\nu_e$), while the  $\nu_\tau$ is  purely left-handed.
At higher orders, $\nu_{\tau L}$ may get a mass with the $\nu_{\tau R}^c$, but
this will only serve
to mix the $\nu_{\tau L}$ with the $\nu_e$.  We note that this model  does not
generate the MSW solution \cite{MSW} to the solar neutrino problem.  We also
note that the extra right-handed neutrinos in this model will pose no problems
for Big Bang nucleosynthesis \cite{KTA}.

The CKM mixing angles among the quarks, and the mixing angles between down-type
quarks and the $m$ quark must arise from new physics at the 1000 TeV scale and
above.  This physics may be related to the interactions that were invoked to
break $SU(4)_{PS}$ and $SU(5)_{ETC}$.
Because of this we cannot yet obtain reliable estimates of mixing angles,
CP violating parameters, and masses of down-type quarks.

Next we turn to the mechanism for CP violation.  It was pointed out earlier
that our model contains the additional\footnote{It may be more realistic to
consider models where there is more than one ``vector" quark.} ``vector" quark,
$m$,  necessary to implement the Nelson-Barr mechanism.  This mechanism can
function if the  theory is  CP conserving
(i.e. $\theta_{ETC} = \theta_{PS} = 0$), and if
CP is spontaneously broken by the appearance of complex phases in the masses
which connect the ordinary down-type quarks with the $m$ quark.
More specifically the ETC breaking dynamics must give rise to a $(d,s,b,m)$
mass matrix of the  form:
\begin{equation}
\begin{array}{cc}
& \begin{array}{lc}  d_L\,s_L\,b_L\,\, & m_L
 \end{array} \\
\begin{array}{c}d_R^c\\s_R^c\\b_R^c\\ m_R^c
\end{array} & \left( \begin{array}{cc}
{\rm real}&\begin{array}{c}
 M_1\\  M_2\\  M_3 \end{array}\\
  0\,\,\,\,  0\,\,\,\,  0\, & {\rm real}
\end{array} \right)
\end{array}~,
\label{massmatrix}
\end{equation}
where at least one of $M_1$, $M_2$, and $M_3$ is complex.
Under these conditions,
CP violating phases will appear in the CKM matrix of the ordinary
fermions, but the
determinant of the
mass matrix is real, so the effective strong CP violating parameter
$\overline{\theta}$ is identically zero at tree level in the
low-energy effective theory.  Furthermore, since the breaking is soft,
higher-order corrections will be finite and small.  In the work of Nelson and
Barr \cite{CP}, the form (\ref{massmatrix}) was arranged by a particular choice
of elementary Higgs fields and couplings.  Whether this form will appear in our
dynamical model at the appropriate breaking scale is not clear.  This will
depend on the details of dynamical breaking at 1000 TeV and above\footnote{For
a discussion of how CP may
be broken dynamically see refs. \cite{Peccei,ELP}.}.

It is worth noting that there is not necessarily a problem with CP domain walls
\cite{Peccei}, if
inflation occurs and the reheating temperature is below the scale where CP is
spontaneously
broken \cite{KolbTurn}.  Since baryogenesis must take place below the
inflationary reheating scale, this scenario  is consistent if baryogenesis
occurs at the electroweak scale\footnote{For a review of electroweak
baryogenesis see ref. \cite{EWB}.}.  We also note that  a TC theory with a
family of technifermions (as ours is) will provide a
first-order electroweak phase transition \cite{PW}  (as opposed to one-doublet
TC models, which have second-order, or extremely weak first-order, phase
transitions \cite{PW,Christ}), and thus  allows for  the possibility of
electroweak baryogenesis.

\section{Conclusions}
We have constructed a potentially realistic (not
obviously wrong) ETC model, that can incorporate many of the right
ingredients: $m_t \gg m_b$, $m_\tau\gg m_{\nu_\tau} $, a family
hierarchy, no bad FCNC's, no visible techniaxion, and CP violation
with no strong CP problem.  We have made estimates for
some of the quark and lepton masses in this model.  New physics is expected in
the form of a light (less than
a few hundred GeV) techni-$\rho$ composed of light technineutrinos.
We stress again that our model contains an attractive tumbling scheme below
1000 TeV.  The phenomenologically desired channel can be the most attractive
when both of the two strong gauge interactions are taken into account.  While
there is thus an understanding of how
dynamical symmetry breaking is achieved through tumbling at lower scales, our
understanding of the breaking at high scales is incomplete.  This
could be a result of our ignorance of strongly coupled, chiral
gauge theories, or it may mean that the model is not complete at the
highest scales.   It remains to be seen whether the model
can survive a more detailed scrutiny, and in particular whether extensions
of the model can
provide quantitative estimates of down-type quark masses and mixing angles.

\noindent\medskip\centerline{\bf Acknowledgments}
\vskip 0.15 truein
We thank S. Chivukula, A. Cohen, M. Dugan, E. Eichten, M. Einhorn, E. Gates, H.
Georgi, B. Holdom, L. Krauss,  K. Lane, A. Nelson, E. Simmons, M. Soldate, R.
Sundrum,
M. White, and S. Willenbrock for helpful discussions.   We especially thank A.
Cohen for a careful reading of the manuscript.
This work was partially supported by the Texas National Laboratory
Research Commission through an SSC fellowship and under contracts \#RGFY93-278
and  \#RGFY93-272; by the Natural Sciences and Engineering
Research Council of Canada; and by the Department of Energy under
contracts \#DE-FG02-92ER40704 and \#DE-FG02-91ER40676.  We thank the Aspen
Center for physics, where part of this work was completed during the summers of
 1991, 1992, and 1993.

\vskip 0.15 truein



\begin{thebibliography}{99}



\bibitem{walking}
B. Holdom, {\em Phys. Lett.} {\bf B150} (1985) 301;
T. Appelquist, D. Karabali, and L.C.R. Wijewardhana, {\em
Phys. Rev.
Lett.} {\em 57} (1986) 957;
K. Yamawaki, M. Bando, and K. Matumoto, {\em Phys. Rev. Lett.}
{\bf 56} (1986) 1335
T. Appelquist and L.C.R. Wijewardhana, {\em Phys. Rev}
{\bf D35}
(1987) 774;
T. Appelquist and L.C.R. Wijewardhana, {\em Phys. Rev}
{\bf D36}
(1987) 568.

\bibitem{strongETC}
T. Appelquist, M. Einhorn, T. Takeuchi, and L.C.R.
Wijewardhana, {\em Phys. Lett.} {\bf 220B}, 223 (1989);
V.A. Miransky and K. Yamawaki, {\em Mod. Phys. Lett.}
{\bf A4}
(1989) 129;
K. Matumoto {\em Prog. Theor. Phys. Lett.} {\bf 81}
(1989) 277;
V.A. Miransky, M. Tanabashi, and K. Yamawaki, {\em Phys.
Lett.}
{\bf B221} (1989) 177;
V.A. Miransky, M. Tanabashi, and K. Yamawaki, {\em Mod. Phys.
Lett.}
{\bf A4} (1989) 1043.

\bibitem{KMEN}M. Einhorn and D. Nash, {\em Nucl. Phys.} {\bf B371} (1992)
32; S. King and S. Mannan, {\em Nucl. Phys.} {\bf B369}
(1992) 119.

\bibitem{Sundrum} R. Sundrum, {\em Nucl. Phys.} {\bf B395} (1993) 60.

\bibitem{revenge}T. Appelquist and J. Terning,  {\em Phys. Lett.} {\bf B315}
(1993) 139.

\bibitem{Randall}S. Dimopoulos, H. Georgi, and S. Raby, {\em Phys. Lett.} {\bf
B127} (1983) 101; S.-C. Chao and K. Lane, {\em Phys. Lett.} {\bf B159} (1985)
135;
R.S. Chivukula and H. Georgi, {\em Phys. Lett.} {\bf B188} (1987) 99;
L. Randall, {\em Nucl. Phys.} {\bf B403} (1993) 122.

\bibitem{EL} E. Eichten and K. Lane, {\em Phys. Lett.}
{\bf B90} (1980) 125.

\bibitem{Ellis} S. Dimopoulos and J. Ellis, {\em Nucl. Phys.} {\bf B182} (1981)
505.

\bibitem{rho}M. Einhorn, D. Jones, and M. Veltman,
{\em Nucl. Phys.} {\bf B191} (1981) 146;
A. Cohen, H. Georgi, and B. Grinstein {\em Nucl. Phys.} {\bf B232}
(1984) 61.

\bibitem{PT}M. Golden and L. Randall, {\em Nucl. Phys.} {\bf B361}, 3 (1991);
B. Holdom and J. Terning, {\em Phys. Lett} {\bf B247}, 88 (1990);
M. Peskin and T. Takeuchi, {\em Phys. Rev. Lett.} {\bf 65}, 964 (1990);
A. Dobado, D. Espriu, and M. Herrero, {\em Phys. Lett.} {\bf B253},
161 (1991);
M. Peskin and T. Takeuchi {\em Phys. Rev.} {\bf D46}
381 (1992), J. Rosner, in {\em The Fermilab Meeting DPF92}, edited by C.
Albright et. al. (World Scientific, Singapore, 1993) 352;
D. Kennedy, Fermilab preprint {\bf FERMILAB-CONF-93/023-T};

\bibitem{CHS}R. Sundrum and S. Hsu, {\em Nucl. Phys.} {\bf B391} (1993) 127;
R. Chivukula, M. Dugan, and M. Golden, {\em Phys. Lett} {\bf B292}
(1992) 435; {\em Phys. Rev.} {\bf D47} (1993) 2930;
T. Appelquist and J. Terning, {\em Phys. Rev.} {\bf D47} (1993) 3075.

\bibitem{HP}  B. Holdom and M. Peskin, {\em Nucl. Phys.}
{\bf 208}
(1982) 397.

\bibitem{PS} J.C. Pati and A. Salam, {\em Phys. Rev. Lett.} {\bf 31} (1973)
661; {\em Phys. Rev.} {\bf D8} (1978) 1240.

\bibitem{GMRS} M. Gell-Mann, P. Ramond, and R. Slansky, in {\em
Supergravity}, ed. P. van Nieuwenhuizen and D. Freedman (North-Holland,
Amsterdam, 1979);  T. Yanagida, {\em Prog. Theo. Phys.} {\bf B135}
(1978) 66.

\bibitem{models}
B. Holdom, {\em Phys. Rev.} {\bf D23} (1981) 1637;
{\em Phys. Lett.} {\bf B246} (1990) 169.

\bibitem{SSVZ} P. Sikivie et. al., {\em Nucl. Phys.}
{\bf B173} (1980) 189.

\bibitem{color}B. Holdom, {\em Phys. Rev. Lett.} {\bf 60} (1988)
1233;

\bibitem{color2}T. Appelquist and O. Shapira, {\em Phys. Lett.}
{\bf B249} (1990) 327.

\bibitem{Georgi} H. Georgi, {\em Nucl. Phys.} {\bf B156} (1979)
126.

\bibitem{CP}A. Nelson, {\em Phys. Lett.} {\bf B136} (1984) 387;
S. Barr, {\em Phys. Rev. Lett.} {\bf 53} (1984) 329;
S. Barr, {\em Phys. Rev.} {\bf D34} (1986) 1567;
K. Choi, D. Kaplan, and A. Nelson, {\em Nucl. Phys.} {\bf B391} (1993) 515.

\bibitem{MAC1}S. Dimopoulos, S. Raby, and L. Susskind, {\em Nucl. Phys.} {\bf
B169}
(1980) 373.

\bibitem{MAC2}J. Cornwall {\em Phys. Rev.} {\bf D10} (1974) 500.

\bibitem{ALM} T. Appelquist, K. Lane, and U. Mahanta,
{\em Phys. Rev. Lett.} {\bf 61} (1988) 1553.

\bibitem{Witten}E. Witten, {\em Phys. Lett.} {\bf B117} (1982) 324.

\bibitem{Zbb}R.S. Chivukula, S.B. Selipsky, and E.H. Simmons, {\em Phys. Rev.
Lett.} {\bf 69} (1992) 575;  E.H. Simmons, R.S. Chivukula, and S.B. Selipsky,
in  {\em Physics Beyond the Standard Model III}, edited by S. Godfrey and P.
Kalyniak (World Scientific, Singapore, 1993) 411;
R.S. Chivukula, E. Gates,  E.H. Simmons, and J. Terning, {\em Phys. Lett.} {\bf
B311} (1993) 157.

\bibitem{vac} M. Peskin, {\em Nucl. Phys.} {\bf B175} (1980) 197;
J. Preskill, {\em Nucl. Phys.} {\bf B177} (1981) 21.

\bibitem{MSW} S. Mikheyev and A. Smirnov, {\em Sov. J. Nucl. Phys.} {\bf 42}
(1985) 913;
L. Wolfenstein, {\em Phys. Rev.} {\bf D17} (1987) 2369.

\bibitem{KTA} L. Krauss, J. Terning, and T. Appelquist,
{\em Phys. Rev. Lett.} {\bf 71} (1993) 823.

\bibitem{Peccei}R. Peccei in {\em Strong Coupling Gauge Theories and Beyond}
edited by T. Muta and K. Yamawaki (World Scientific, Singapore, 1991) 134.

\bibitem{KolbTurn} E. Kolb and M. Turner, {\em The Early Universe}
(Addison-Wesley, Redwood City, 1990).

\bibitem{EWB} A. Cohen, D. Kaplan, and A. Nelson,  U.C. San Diego preprint {\bf
 UCSD-93-2},  {\bf BU-HEP-93-4}, {\bf hep-ph/9302210}, to appear in {\em Annual
Review of Nuclear and Particle Science} {\bf 43}.

\bibitem{PW}R. Pisarski and F. Wilczek, {\em Phys. Rev.}  {\bf D29} (1984) 338.

\bibitem{Christ} F. Brown et. al. {\em Phys. Rev. Lett.} {\bf 65} (1990) 2491.

\bibitem{ELP} E. Eichten, K. Lane, and J. Preskill, {\em Phys. Rev. Lett.} {\bf
45} (1980) 225;
K. Lane {\em Physica Scripta} {\bf 23} (1981) 1005.

\end{thebibliography}
\end{document}